\documentclass[aps,prb,reprint,superscriptaddress]{revtex4-1}
\usepackage{amsmath}
\usepackage{multirow}
\usepackage{graphicx}
\usepackage{epstopdf}
\usepackage{tabularx, booktabs}
\usepackage{pbox}
\newcolumntype{Y}{>{\centering\arraybackslash}X}
\usepackage[usenames,dvipsnames]{color}
\usepackage[caption=false,position=b,singlelinecheck=off,font=normalsize,labelfont=bf,justification=justified]{subfig}
\usepackage{hyperref}
\hypersetup{
   	colorlinks=true,
	linkcolor=blue,
	filecolor=black,
	urlcolor=blue,
	citecolor=blue,
}

\graphicspath{ {Figures/} }

\usepackage{xr}
\begin{document}

\title{Possible quantum fluctuations in the vicinity of the quantum critical point of $\mathbf{(Sr, Ca)_3Ir_4Sn_{13}}$ revealed by high-energy X-ray diffraction study  }

\author{L.S.I. Veiga}

\affiliation{Deutsches Elektronen-Synchrotron (DESY), Hamburg 22607, Germany}

\affiliation{London Centre for Nanotechnology and Department of Physics and Astronomy, University College London, Gower Street, London, WC1E 6BT, United Kingdom}

\author{J.R.L. Mardegan}

\affiliation{Deutsches Elektronen-Synchrotron (DESY), Hamburg 22607, Germany}

\author{M.v. Zimmermann}

\affiliation{Deutsches Elektronen-Synchrotron (DESY), Hamburg 22607, Germany}

\author{D.T. Maimone}
\affiliation{Instituto de F\'isica "Gleb Wataghin", Universidade Estadual de Campinas-UNICAMP, Campinas, S\~ao Paulo 13083-859, Brazil}

\author{F.B. Carneiro}
\affiliation{Instituto de F\'isica, Universidade do Estado do Rio de Janeiro, 20550-900, Rio de Janeiro, RJ, Brazil}
\affiliation{Centro Brasileiro de Pesquisas F\'isica, 22290-180, Rio de Janeiro, RJ, Brazil}

\author{M.B. Fontes}
\affiliation{Centro Brasileiro de Pesquisas F\'isica, 22290-180, Rio de Janeiro, RJ, Brazil}

\author{J. Strempfer}
\affiliation{Deutsches Elektronen-Synchrotron (DESY), Hamburg 22607, Germany}

\author{E. Granado}
\affiliation{Instituto de F\'isica "Gleb Wataghin", Universidade Estadual de Campinas-UNICAMP, Campinas, S\~ao Paulo 13083-859, Brazil}

\author{P.G. Pagliuso}
\affiliation{Instituto de F\'isica "Gleb Wataghin", Universidade Estadual de Campinas-UNICAMP, Campinas, S\~ao Paulo 13083-859, Brazil}

\author{E.M. Bittar}
\affiliation{Centro Brasileiro de Pesquisas F\'isica, 22290-180, Rio de Janeiro, RJ, Brazil}

\date{\today{}}
\begin{abstract}
We explore the evolution of the structural phase transition of $\rm{(Sr, Ca)_3Ir_4Sn_{13}}$, a model system to study the interplay between structural quantum criticality and superconductivity, by means of high-energy x-ray diffraction measurements at high pressures and low temperatures. Our results confirm a rapid suppression of the superlattice transition temperature $T^*$ against pressure, which extrapolates to zero at a critical pressure of $\approx 1.79(4)$ GPa. The temperature evolution of the superlattice Bragg peak in $\rm{Ca_3Ir_4Sn_{13}}$ reveals a drastic decrease of the intensity and an increase of the linewidth when $T \rightarrow 0$ K and $p \rightarrow p_c$. Such anomaly is likely associated to the emergence of quantum fluctuations that disrupt the formation of long-range superlattice modulation. The revisited temperature-pressure phase diagram of $\rm{(Sr, Ca)_3Ir_4Sn_{13}}$ thus highlights the intertwined nature of the distinct order parameters present in this system and demonstrates some similarities between this family and the unconventional superconductors.


\end{abstract}

\maketitle

\section{Introduction}

Compounds displaying the interplay between superconductivity (SC) and electronic instabilities have been extensively studied during the past years due to their rich phase diagrams as a function of doping, pressure or magnetic fields\cite{Fradkin2015,Scalapino2012,Paglione2010,Keimer2015,Gegenwart2008,Gruner2017}. In most cases, SC is found in the vicinity of electronic instabilities of magnetic origin, where the pairing mechanism is mediated by spin fluctuations and the SC is unconventional\cite{Monthoux2007,Mathur1998,Slooten2009}. The proximity of SC to nonmagnetic structural instabilities, on the other hand, is rare and searches for a quantum critical point (QCP) resulting from a tunable structural phase transition has provoked great interest due to its promising role of stabilizing or even enhancing the pairing mechanism. Thus, accessing SC materials where a detailed study of structural quantum criticality and its impact on SC can be explored is highly desirable.

In this context, the ternary intermetallic stannides such as R$_3$T$_4$Sn$_{13}$, where R=Sr, Ca and T=Ir, Rh\cite{Remeika1980,Espinosa1980}, have attracted special attention due to the existence of a second-order structural phase transition below $T^*$, its putative coexistence with a SC state below $T_C$ and its suppression upon applying pressure or chemical substitution\cite{Kase2011, Zhou2012, Mazzone15, Klintberg12, Biswas15, Yu2015, Cheung2018, Goh2015, Lue2016, Cheung2017}. Conventional phonon-mediated BCS character of the superconductivity with s-wave symmetry has been established and confirmed in these systems by a range of experimental probes~\cite{Biswas15, Gerber2013, Biswas2014, Biswas2014_2, Kase2011, Hayamizu2010}. However, their resulting phase diagram is very suggestive of strong interplay between different order parameters and remarkably resembles the phase diagrams of the heavily studied unconventional SC\cite{Paglione2010,Gegenwart2008,Hashimoto2012,Shibauchi2014,Luo2018}. The role of these order parameters and whether they are coexisting, competing or cooperating between each other are still a matter of debate.

Sr$_3$Ir$_4$Sn$_{13}$ and Ca$_3$Ir$_4$Sn$_{13}$ exhibit an anomaly in the temperature-dependent electrical resistivity and magnetic susceptibility measurements below $T^*\sim147$ K and $T^*\sim33$ K, followed by a SC transition at $T_C= 5$ K and $T_C=7$ K, respectively\cite{Klintberg12}. Single-crystal x-ray diffraction and neutron-scattering measurements revealed that such anomaly is produced by a second-order structural phase transition from a simple cubic parent structure (space group $Pm\overline{3}n$ at 300 K), to a superlattice variant ($I\overline{4}3d$), where the lattice parameter is twice that of the room temperature phase\cite{Klintberg12, Mazzone15}. The complete substitution of Sr by Ca, which corresponds to a positive pressure of $\sim5$ GPa, reduces $T^*$ and this behavior continues for the (Ca$_{1-x}$Sr$_x$)$_3$Ir$_4$Sn$_{13}$ series under external pressure. Full suppression of $T^*$ is predicted at a structural quantum critical point $\approx1.8$ GPa for Ca$_3$Ir$_4$Sn$_{13}$\cite{Klintberg12, Biswas15}. Several experimental probes suggest that such structural instability is associated with a charge density wave (CDW) transition involving the conduction electrons\cite{Klintberg12}. Such idea is supported by a decrease in the carrier density and a formation of a partial energy gap at the Fermi surface through the onset of the structural phase transition\cite{Kuo2014, Wang2015, Wang2012, Fang2014, Chen2015}. Moreover, muon spin relaxation measurements revealed a strong enhancement of the superfluid density and a dramatic increase of the pairing strength above $\approx1.6$ GPa, giving evidence of the presence of a QCP\cite{Biswas15}. Although several investigations have been realized on this system, in none of these experiments has the CDW been microscopically probed inside the SC phase and the question remains whether it survives in this low temperature phase. Since the feature associated with the structural transition is weakened when approaching the QCP, a range of experimental probes have so far failed to identify it. In this sense, whether the CDW and SC states coexist and the exact pressure at which the CDW disappears are yet to be determined or confirmed.

With the improvement of x-ray diffraction (XRD) techniques to an extended pressure range it is now possible to explore the evolution of CDW modulation and their instabilities when approaching a pressure-driven QCP\cite{vonZimmerman08, Huecker10, Chang2012}. X-ray diffraction is a particularly valuable technique since it provides direct microscopic insight into the CDW modulation, allowing the determination of the CDW wave vector and the temperature and/or magnetic field dependences of the order parameter and correlation lengths. Thus, its combination with high pressure instrumentation provides a powerful tool for manipulating the nature of charge order in emergent materials.

Here, using high-energy x-ray diffraction measurements we have performed a detailed study of the evolution of the superlattice structure of $\rm{(Sr, Ca)_3Ir_4Sn_{13}}$ against pressure and temperature. We find that the superlattice transition temperature $T^*$ is rapidly suppressed with increasing pressure and extrapolates to zero at a critical pressure of $\approx 1.79(4)$ GPa, in agreement with previous studies~\cite{Biswas15, Klintberg12}. Our XRD measurements on ${\rm Ca_3Ir_4Sn_{13}}$ revealed an anomaly related to a partial suppression of the superlattice peak intensity, which takes place at low temperatures ($T<15$ K) and under pressures ($p>0.09$ GPa). Such anomaly is also manifested by a large decrease of the static coherence length ($\xi$) when the temperature approaches to zero. Since information about fluctuations can also be obtained from the Bragg diffraction peaks coming from a static order parameter, our results suggest that quantum fluctuation effects is likely the mechanism behind the destruction of the long-range CDW modulation in ${\rm Ca_3Ir_4Sn_{13}}$. The presence of strong quantum fluctuations competing with CDW modulation and possibly with SC makes the phase diagram of  $\rm{(Sr, Ca)_3Ir_4Sn_{13}}$ reminiscent of unconventional SC.

\section{Samples and Methods}
Single crystals of ${\rm (Sr,Ca)_3Ir_4Sn_{13}}$ were grown by flux method as described elsewhere\cite{Israel2005}. The crystal structure and phase purity were determined by XRD on powdered crystals (not shown). Ambient pressure synchrotron XRD data ($E=8.33$ keV) were collected on single crystals ($\approx 2 \times 1 \times 1$ $\rm{mm^3}$) at beamline P09 at PETRA III, DESY\cite{Strempfer2013}. The high quality of the crystal was verified by a mosaic spread of $0.01^\circ$ determined at the $(0,4,1)$ Bragg reflection in the room temperature $Pm\overline{3}n$ phase.

High pressure single crystal XRD experiment on $\rm{Ca_3Ir_4Sn_{13}}$ was performed at P07 beamline of PETRA III, DESY. A single crystal of $\sim 1 \times 1 \times 0.5$ $\rm{mm^3}$ dimension was cut and polished to achieve a flat and shiny surface perpendicular to the $[001]$ direction. The measurements were performed using a clamp-type pressure cell\cite{vonZimmerman08} inserted in a 10 T cryomagnet installed on top of the triple-axis diffractometer. Pressure calibration was determined by measuring the pressure-dependence of the orthorhombic splitting of $(2,0,0)$/$(0,2,0)$ Bragg peaks on ${\rm La_{1.875}Ba_{0.125}CuO_4}$ as described in Refs~\onlinecite{vonZimmerman08, Huecker10}. Further details of the pressure calibration can be found in the Supplemental Material \footnote{See Supplemental Material at [URL will be inserted by publisher] for further information.}. Daphne oil was used as pressure transmitting fluid. The bulk properties of the superlattice modulation as well as the crystal structure were studied by transmission geometry taking advantage of the large penetration depth and wide range of reciprocal space allowed by the high energy photons ($E=98.7$ keV). All crystal directions and scattering vectors, $\bf{Q}$=$(h, k, l)$, are specified in units of ($2\pi/a, 2\pi/b, 2\pi/c$) of the room temperature cubic unit cell $Pm\overline{3}n$ ($a=b=c=9.72$ \AA). Access to the $(h,k,l)$ Bragg and superlattice peaks was obtained by aligning either the $a-c$ axes horizontally, with the $c$ axis approximately along the magnetic field and beam direction. The initial alignment of the single crystal was performed by collecting the diffraction patterns with a PerkinElmer detector; once the superlattice peaks were identified, a point detector was used (consisting of a broad band analyzer crystal and a scintillation counter).

\begin{figure}
	\centering
	\includegraphics[width=1\columnwidth]{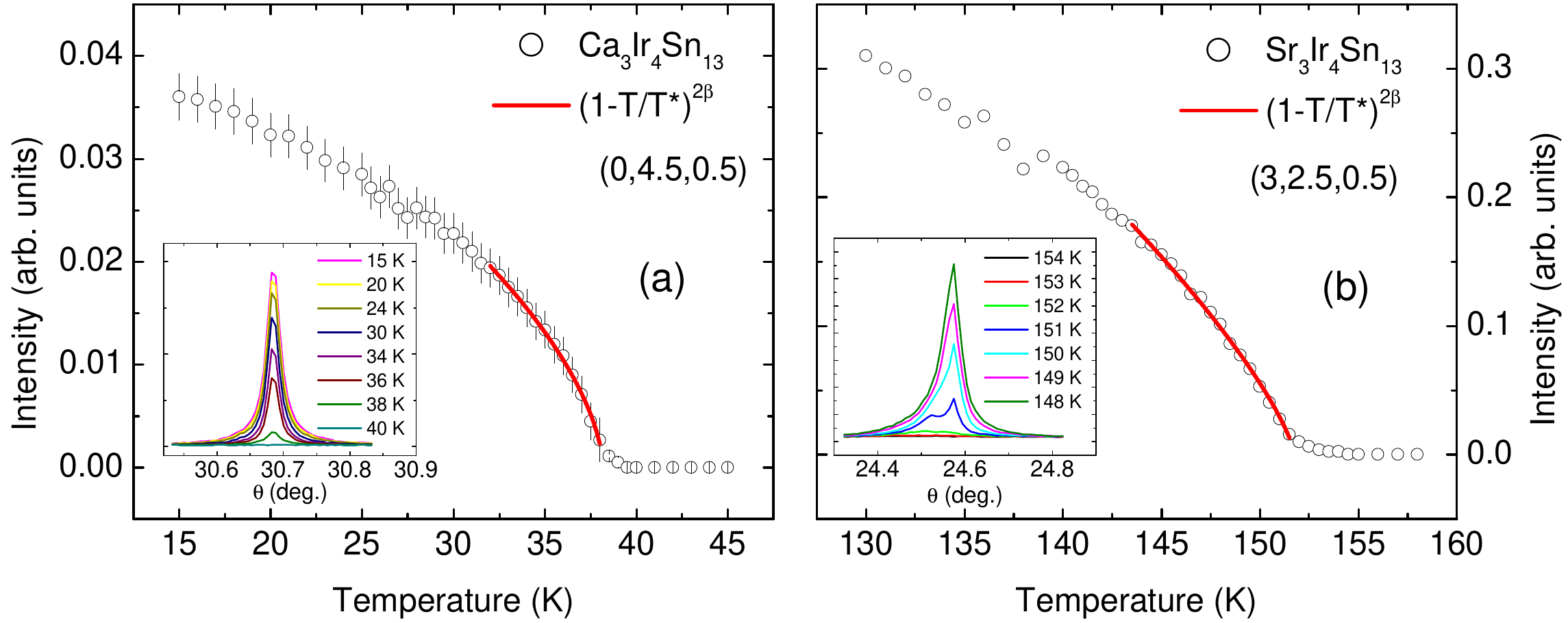}

	\caption{ Temperature dependence of the $(0, 4.5, 0.5)$ and $(3, 2.5, 0.5)$ superlattice reflections at ambient pressure for (a) Ca$_3$Ir$_4$Sn$_{13}$ and (b) Sr$_3$Ir$_4$Sn$_{13}$, respectively. The red solid line is a fitting using a power law to determine $T^*$. The insets show the rocking curves around the superlattice reflection measured for selected temperatures for each compound.} 
	\label{fig:FigS1}

\end{figure}

High pressure single crystal XRD experiment on $\rm{Sr_3Ir_4Sn_{13}}$ was performed at XDS beamline of the Brazilian Synchrotron Light Source (LNLS)\cite{Lima2016}. The diamond anvil cell (DAC) was placed in the coldfinger of a He closed-cycle cryostat and data was collected using a Pilatus 300K detector. Due to the DAC limited angular scattering range (25$^\circ$ of scattering angle $2\theta$), the beam was tuned to $E=20$ keV in order to detect a significant number of Bragg peaks. The DAC was prepared with two full diamonds of 600 $\rm{\mu m}$ culet size and a single crystal of $\sim 80\times80\times40$ $\rm{\mu m^3}$ was loaded together with ruby crystals for \emph{in-situ} pressure calibration and 4:1 methanol:ethanol as pressure media. The single crystal XRD measurements were performed in transmission geometry, vertical scattering, with $c$ axis along the beam direction.

\section{Results}

\subsection{Ambient pressure XRD measurements}

At ambient pressure, single crystal XRD measurements on $\rm{Ca_3Ir_4Sn_{13}}$ and Sr$_3$Ir$_4$Sn$_{13}$ reveal a series of satellite peaks below $T^*=38.2(1)$ K and $T^*=151.2(1)$ K, respectively, at $\mathbf{Q_{\rm SL}}= \mathbf{\tau+q_{\rm SL}}$, where $\tau$ is the wave vector of the room temperature phase and $\mathbf{q_{\rm SL}}=(0.5, 0.5, 0)$ the propagation vector of the superlattice structure. No reflections associated with $\mathbf{q_{\rm SL}}=(0.5,0.5,0.5)$ or $(0,0,0.5)$ modulations were found, in agreement with previous studies \cite{Mazzone15, Tompsett14}. The propagation vector $\mathbf{q_{\rm SL}}=(0.5, 0.5, 0)$ seems to be the benchmark in these 3-4-13 family of compounds, as it was also confirmed in $\rm{(Sr,Ca)_3Rh_4Sn_{13}}$  \cite{Cheung2018}, $\rm{La_3Co_4Sn_{13}}$ \cite{Ferreira2019} and even in Eu$_3$Ir$_4$Sn$_{13}$~\cite{Mardegan2013}, which also displays magnetic ordering at low temperatures. Figure~\hyperref[fig:FigS1]{1} shows the temperature dependence of the integrated intensity for the superlattice reflections $(0,4.5,0.5)$ and $(3, 2.5, 0.5)$ for Ca$_3$Ir$_4$Sn$_{13}$ and Sr$_3$Ir$_4$Sn$_{13}$, respectively. A continuous decrease of the superlattice peak indicates a second-order phase transition at $T^*$. The temperature dependent data was fitted by a power-law expression $\propto (1-T/T^*)^{2\beta}$ yielding a critical exponent of $\beta=0.30(1)$ for Ca$_3$Ir$_4$Sn$_{13}$ and $\beta=0.29(1)$ for Sr$_3$Ir$_4$Sn$_{13}$, characteristics of a three-dimensional character transition.

\begin{figure*}[t]
	\centering
	\includegraphics[width=2 \columnwidth]{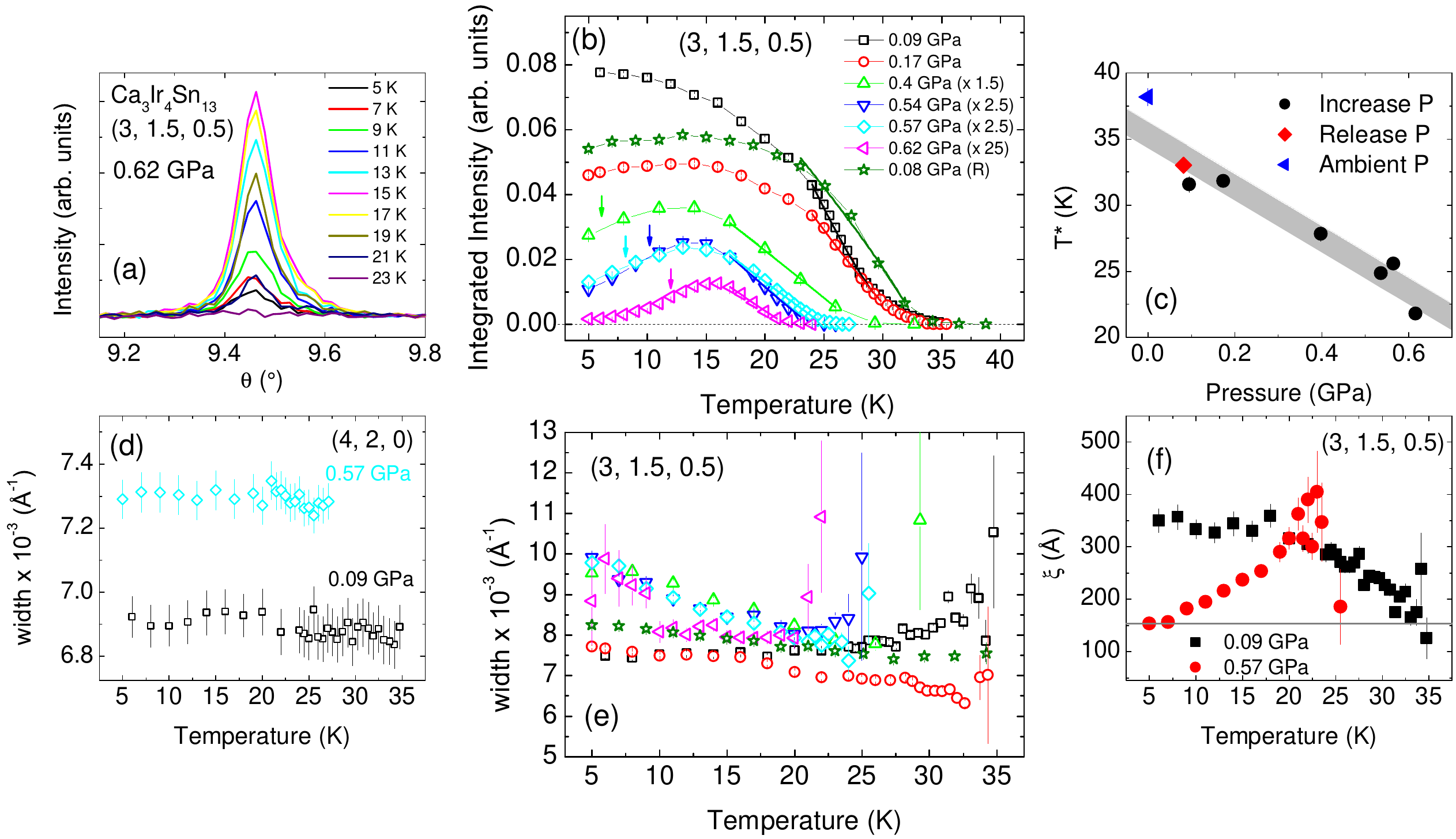}

	\caption{(Color online) (a) Evolution of the $(3,1.5,0.5)$ superlattice peak intensity at $p=0.62$ GPa. (b) Temperature dependence of the $(3,1.5,0.5)$ superlattice peak intensity at several pressures. Thin solid lines are guides to the eye. Thick solid lines are the fittings to the power law $\propto (1-T/T^*)^{2\beta}$. The arrows indicates the value of $T'$, extracted from the maximum of the first derivative of the superlattice peak intensity with respect to temperature (see Ref.~\onlinecite{Note1} for further details). The pressure dependence of $T^*$ for ${\rm Ca_3Ir_4Sn_{13}}$ is shown in (c). (d, e) Evolution of the pseudo-Voigt linewidth against temperature at selected pressures for $(4,2,0)$ Bragg and $(3,1.5,0.5)$ superlattice peaks, respectively, extracted from 2$\theta$ scans. (f) Temperature dependence of the static correlation length, $\xi$, of the $(3,1.5,0.5)$ superlattice reflection at $p=0.09$ GPa and $p=0.57$ GPa. The grey line indicates the similar correlation lengths observed at $T=5$ K and $p=0.57$ GPa and at $T\sim32$ K and $p=0.09$ GPa. Details of the calculation of the correlation length can be found in Ref. \onlinecite{Note1}. The estimated pressure error bar is $\pm 0.1$ GPa.}
	\label{fig:Fig1}

\end{figure*}

\subsection{High-pressure XRD measurements on $\bf{Ca_3Ir_4Sn_{13}}$}

 The temperature and pressure dependence of the  superlattice peak $\mathbf{Q_{\rm SL}}=(3,1.5,0.5)$ are summarized in Figure~\hyperref[fig:Fig1]{2(a-f)}. The measurements consisted of rocking (Fig.~\hyperref[fig:Fig1]{2(a)}) and $2\theta$ scans collected at several pressures and temperatures. At $p=0.09$ GPa, the integrated intensity of the superlattice peak grows gradually on cooling below $\sim32$ K (see Figure~\hyperref[fig:Fig1]{2(b)}). A saturation of the peak intensity seems to take place below the superconducting temperature $T_C \sim7$ K, as also observed at ambient pressure. Interestingly, for pressures higher than $p\sim0.17$ GPa, the superlattice peak intensity is partially suppressed below $T\sim15$ K. This suppression is enhanced upon pressure increase up to $p=0.62$ GPa, above which total suppression of the superlattice peak intensity is observed (see Fig. S1 in the Supplemental Material \cite{Note1} for the $p=0.7$ GPa data set). The temperature dependence of the superlattice peak intensity for different pressures (Fig.~\hyperref[fig:Fig1]{2(b)}) were fitted by a power law $\propto (1-T/T^*)^{2\beta}$ and the best fit to the data near $T^*$ corresponds to the critical temperatures displayed in Figure~\hyperref[fig:Fig1]{2(c)}. 

Further insight into the partial suppression of the superlattice peak intensity is given by the temperature dependence of the pseudo-Voigt linewidths. The linewidths were extracted from the 2$\theta$ scans at the lattice and superlattice Bragg peaks $(4,2,0)$ and $(3,1.5,0.5)$, respectively (Figs.~\hyperref[fig:Fig1]{2(d)} and \hyperref[fig:Fig1]{2(e)}, respectively). At ambient pressure, the superlattice peak is resolution limited, while it develops a small but finite width at 0.09 GPa (correlation length of $\xi \sim 340$ {\AA}) indicating the CDW is long-range ordered (see Supplemental Material \cite{Note1}). At low pressures ($p \lesssim0.17$ GPa), the linewidth of the superlattice modulation (Fig.~\hyperref[fig:Fig1]{2(e)}) is comparable to that of the Bragg reflection (Fig.~\hyperref[fig:Fig1]{2(d)}) and is mostly temperature-independent at low temperatures. Upon further pressure increase, the linewidth increases 30\% from $p=0.09$ GPa to $p=0.57$ GPa at $T=5$ K. Surprisingly, the widths at low temperatures ($T \sim 5$ K, $p=0.54-0.62$ GPa) are comparable to the values observed in proximity to the structural phase transition ($T^*\sim 20-32$ K, $p=0.09-0.62$ GPa), indicating that a competing order of similar energy scales is likely to be developing at low temperatures and high pressures. This same feature is highlighted in Fig.~\hyperref[fig:Fig1]{2(f)}, where the temperature dependence of the correlation length of the superlattice peak $(3,1.5,0.5)$ for two different pressures are plotted: a correlation length of $\xi \sim 153$ {\AA} can be observed in either $T=5$ K and $p=0.57$ GPa or $p=0.09$ GPa and $T\sim32$ K curves.

\begin{figure}[t]
	\centering
	\includegraphics[width=0.9 \columnwidth]{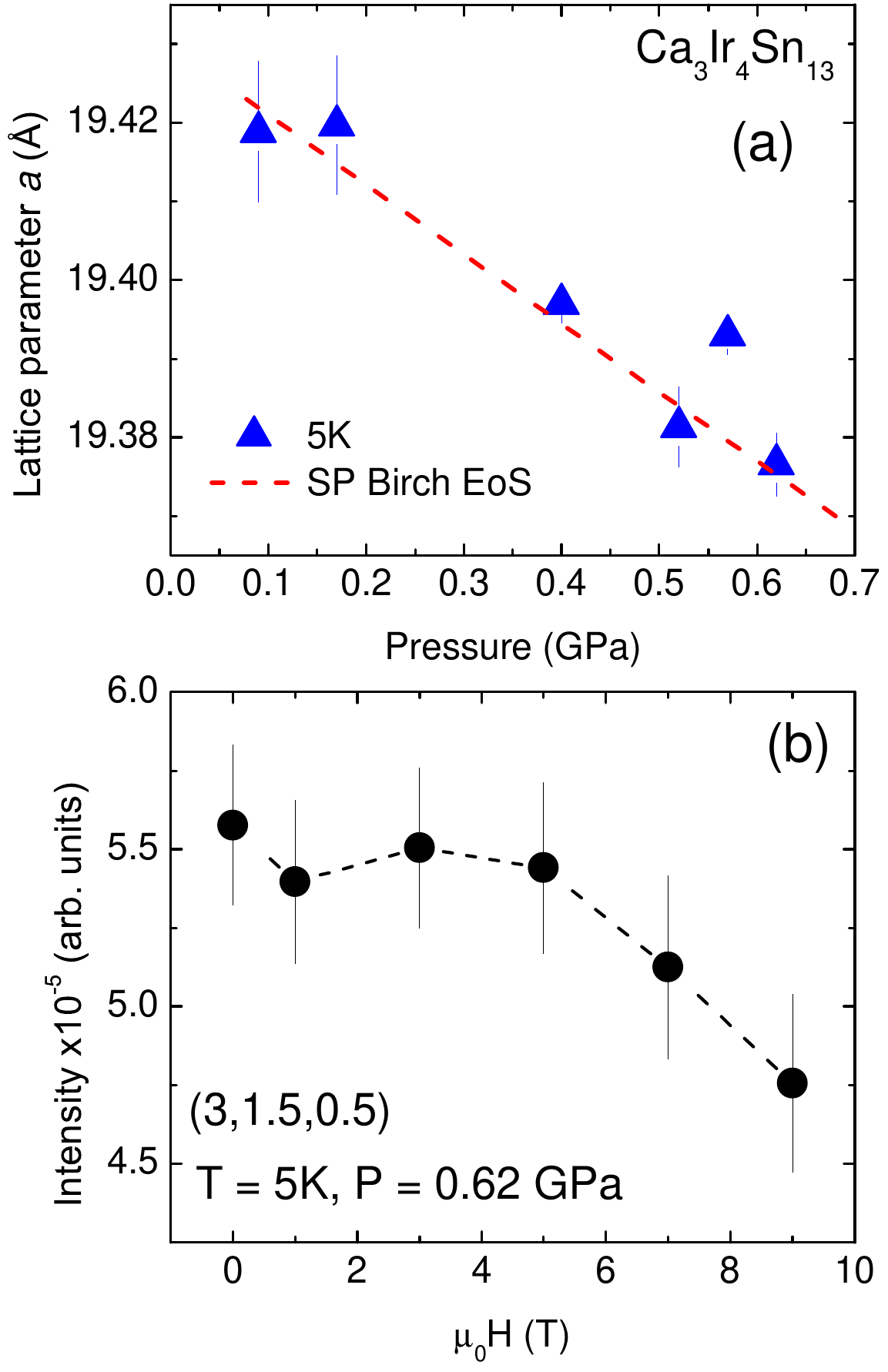}

	\caption{(Color online) (a) Pressure dependence of the lattice parameter {\em a} extracted from lattice Bragg reflections in the low temperature $I\overline{4}3d$ space group and its fit to a single parameter Birch equation of state (EoS). (b) Magnetic field dependence of the superlattice modulation peak intensity at $(3, 1.5, 0,5)$ for $T=5$ K and $p=0.62$ GPa. The dataset was collected with a magnetic field applied along $\mathbf{Q_{\rm SL}}=(3, 1.5, 0.5)$ direction.}
	\label{fig:Fig2}

\end{figure}

We have also explored the crystal structure of $\rm{Ca_3Ir_4Sn_{13}}$ under pressure. The evolution of the cubic lattice parameter \emph{a} at $T=5$ K was obtained through the analysis of selected structural Bragg peaks ($(0,0,4)$ and $(4,2,0)$). Within our experimental accuracy, no discontinuities in lattice parameters or signatures of a structural phase transition were found in the entire pressure range measured (Fig.~\hyperref[fig:Fig2]{3(a)}). Such lattice constant is well-characterized by a single-parameter Birch equation of state (EoS) with bulk modulus of $B=72(13)$ GPa and \emph{a}-axis compression rate of $\frac{\Delta a / a_0}{\Delta P}= -0.4(1)$ $\%$/GPa, where $a_0$ is the lattice parameter at $p=0.09$ GPa.


In order to verify whether the partial suppression of the superlattice intensity is due to a competition between CDW and superconductivity, we have probed the effect of application of magnetic field on the superlattice modulation. A maximum field of 9 T was applied along the $\mathbf{Q_{\rm SL}}=(3, 1.5, 0.5)$ direction. The field dependence of the intensity of the superlattice reflection $(3,1.5,0.5)$ at $T=5$ K is shown in Fig.~\hyperref[fig:Fig2]{3(b)}. Application of a magnetic field has no significant effect, considering the experimental errors, on the superlattice peak intensity. We note that previous studies~\cite{Zhou2012, Goh2011} report that the upper critical field is $H_c\approx 7$ T and does not vary much under pressure, so a magnetic field of $\mu_0H=9$ T is enough to suppress the superconductivity in this material at $p=0.66$ GPa, where the normal state is disclosed down to 2 K.

\begin{figure}[t]
	\centering
	\includegraphics[width=0.9 \columnwidth]{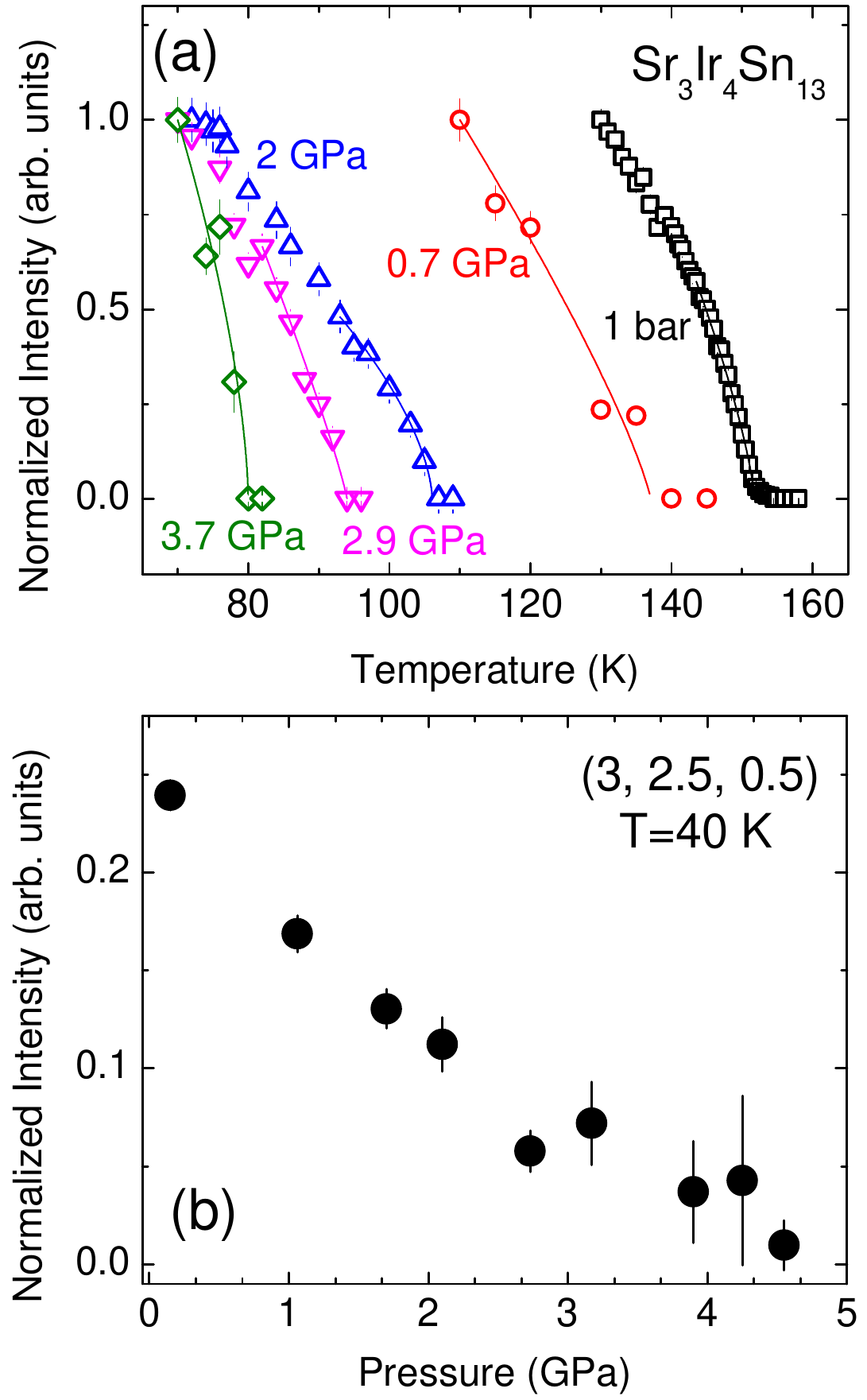}

	\caption{(Color online) (a) Temperature dependence of several superlattice peaks $((3,1.5,-0.5), (3, 2.5, 0.5)$ and $(2.5, 1, -0.5))$ of ${\rm Sr_3Ir_4Sn_{13}}$ at selected pressures. The dataset was normalized to one due to intensity differences among the reflections probed. Solid lines are the fittings to the power law $\propto (1-T/T^*)^{2\beta}$. (b) Pressure dependence of the (3, 2.5, 0.5) superlattice peak integrated intensity at 40 K, which was normalized by the (3, 2, 0) Bragg reflection.}
	\label{fig:Fig3}

\end{figure}

\subsection{High-pressure XRD measurements on $\bf{Sr_3Ir_4Sn_{13}}$}

A detailed single crystal XRD study under pressure was conducted on $\rm{Sr_3Ir_4Sn_{13}}$ compound. Figure~\hyperref[fig:Fig3]{4(a)} displays the temperature dependence of the intensity at several superlattice peaks under pressure. As expected from the high-pressure electrical resistivity measurements~\cite{Klintberg12}, a drastic suppression of the superlattice transition temperature is observed. The temperature dependence of the superlattice peak intensity were also fitted by the power law $\propto (1-T/T^*)^{2\beta}$ and the best fit to the data near $T^*$ corresponds to the critical temperatures displayed in the phase diagram of Fig.~\hyperref[fig:Fig4]{5}. Fig.~\hyperref[fig:Fig3]{4(b)} shows the evolution of the intensity of the superlattice peak $(3,2.5,0.5)$ normalized by the Bragg reflection $(3,2,0)$ against pressure. Due to temperature constraints ($T_{min} \sim 40$ K), the superlattice peak intensity was probed up to the highest pressure of $\sim 4.6$ GPa. The total superlattice peak suppression is expected to take place at $\approx7.7$ GPa based on the linear extrapolation of the pressure evolution of $T^*$, as it will be discussed below.


\section{Discussion}

It has been shown that the combination of physical and chemical pressure has strong influence on the superlattice phase of the 3-4-13 series of compounds, such as the ${\rm (Ca_xSr_{1-x})_3Ir_4Sn_{13}}$, where bulk measurements reveal a suppression of the second-order structural phase transition at $T^*$\cite{Klintberg12, Mazzone15, Biswas15}. Our XRD measurements on $\rm{(Sr, Ca)_3Ir_4Sn_{13}}$ constitute a detailed structural study of this class of material under pressure, contributing to the advancement of the temperature-pressure phase diagram, which up to now are based solely on electrical resistivity, magnetic susceptibility, muon spin relaxation and nuclear magnetic resonance measurements~\cite{Biswas15, Klintberg12, Luo2018}.

\begin{figure}[t!]
	\centering
	\includegraphics[width=1 \columnwidth]{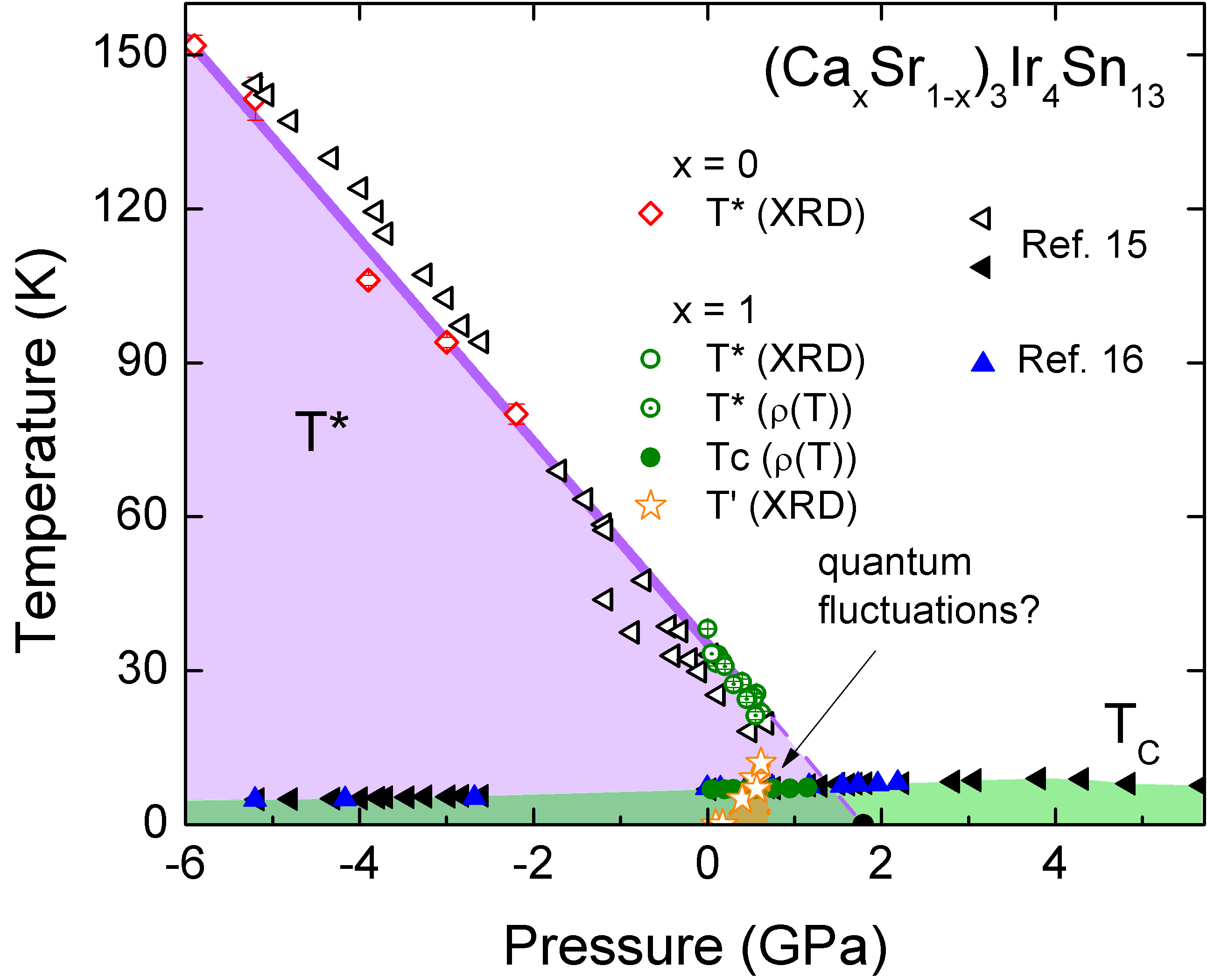}

	\caption{(Color online) Phase diagram of the ${\rm (Ca_xSr_{1-x})_3Ir_4Sn_{13}}$ system proposed by the present high-energy and high-pressure XRD measurements. Based on our data (open diamonds and circles for the XRD data and open circles with dot for the electrical resistivity), the superlattice transition temperature, $T^*$, extrapolates to zero at $\sim 1.79(4)$ GPa. Closed green circles are ${T_C}$ values determined from resistivity measurements\cite{Note1}. Star symbols ($T'$) are the temperatures where the quantum fluctuations emerge. The open and closed left black triangles are the $T^*$ and $T_C$ values, respectively, from resistivity measurements of Refs.~\onlinecite{Klintberg12}. Blue triangles are $T_C$ values obtained from $\mu$SR measurements from Ref.~\onlinecite{Biswas15}.}
	\label{fig:Fig4}

\end{figure}

Among the ${\rm (Ca_xSr_{1-x})_3Ir_4Sn_{13}}$ series at ambient pressure, ${\rm Ca_3Ir_4Sn_{13}}$ displays the smallest difference between its order parameters ($T^* \sim 38$ K and $T_C \sim 7$ K at ambient pressure), thus more prone to external stimuli, such as applied physical pressure. Indeed, our high-energy, high-pressure XRD measurements on ${\rm Ca_3Ir_4Sn_{13}}$ reveal that $T^*$ is rapidly suppressed by pressure at a rate of $dT^*/dP \approx -19.3 \pm 0.3$ K/GPa, with the superlattice modulation intensity vanishing completely above $\sim 0.62$ GPa. Such result corroborates with our resistivity measurements, where the resistivity anomaly associated with $T^*$ was last seen at $T^* \sim 21(1)$ K and $p=0.55(5)$ GPa. Combined with the high-pressure XRD data on ${\rm Sr_3Ir_4Sn_{13}}$, $T^*$ extrapolates to zero yields at a critical pressure of $p_c \sim 1.79(4)$ GPa (black filled half-circle in Fig.~\ref{fig:Fig4}), in striking agreement with values found in literature~\cite{Biswas15, Klintberg12}.

Interesting to note is the pressure-induced partial suppression of the superlattice peak intensity for temperatures below 15 K (Fig.~\hyperref[fig:Fig1]{2(b)}). Such anomaly is also manifested in the pseudo-Voigt linewidth as a function of temperature for selected pressures (Fig.~\hyperref[fig:Fig1]{2(e)}). Closer to ambient pressure, the linewidth has little to no dependence with temperature. For $p>0.09$ GPa and $T<15$ K, the CDW modulation becomes less long-range ordered, with $\mathbf{Q_{SL}}$-width increasing when temperature is lowered to $T\sim5$ K. Fig.~\hyperref[fig:Fig1]{2(f)} shows the correlation length of the CDW modulation, which decreases from $\xi \sim341$ {\AA} at $p=0.09$ GPa to $153$ {\AA} at $p=0.62$ GPa at $T\sim5$ K, a reduction of 55\% (going from $\sim17.6$ to $\sim8$ CDW wavelengths~\footnote{The lattice parameter used here corresponds to the body-centered space group, where lattice constants are double of the room temperature cubic unit cell $Pm\overline{3}n$. Value extracted from evolution of the lattice parameters against pressure at $T=5$ K}). Our results strongly suggests that a competing order of similar energy scales is developing at low temperatures and high pressures. The possible nature of such order parameter and its implications will be discussed below.

Our XRD investigation on the evolution of the lattice parameter against pressure reveals no sign of a structural phase transition within our experimental accuracy, suggesting that the crystal structure remains within the superlattice variant unit cell $I\overline{4}3d$. This was further supported by the pseudo-Voigt linewidths extracted from the lattice Bragg peaks: a 5\% peak broadening was found from $p=0.09$ to 0.57 GPa, but such value is likely attributed to the natural peak broadening related to application of pressure rather than to the onset of a structural phase transition. From the field-dependent data shown in Fig.~\hyperref[fig:Fig2]{3(b)}, we found that an applied magnetic field of 9 T, at $T=5$ K, along $\mathbf{Q_{\rm SL}}=(3, 1.5, 0.5)$, which should be enough to suppress the superconducting phase in ${\rm Ca_3Ir_4Sn_{13}}$~\cite{Goh2011}, has no significant effect on the superlattice modulation intensity. This is in contrast with observations in cuprate materials, such as YBa$_2$Cu$_3$O$_{6.67}$\cite{Chang2012}, where application of magnetic field suppresses superconductivity and enhances the spontaneous CDW ordering with wave vector $\mathbf{q_{\rm{CDW}}}=\mathbf{q_1}=(\delta_1, 0, 0.5)$ and $\mathbf{q_2}=(0, \delta_2, 0.5)$, with $\delta_1=0.3045(2)$ and $\delta_2=0.3146(7)$.





It should be noted that fluctuation effects\cite{Feng2012, Gruner2017, Feng2015} might be playing a crucial role in the decrease in coherence length of the CDW modulation at $T<15$ K and $p>0.09$ GPa. Analysis of the linewidth of the 2$\theta$ scans indicate that at $p=0.09$ GPa the profile shape is mostly Gaussian and evolves towards Lorentzian when approaching the superlattice phase transition at $T^*$ (see Figs. S4 and S5 in Supplemental Material~\cite{Note1}). For higher pressures, on the other hand, the profile shape has a significant contribution from the Lorentzian line shape, indicating that the superlattice phase correlation is exponentially decaying in real space already at low temperatures\cite{Feng2012, Feng2015}. This result could be consistent with a disorder pinning scenario~\cite{Hill1995, Fukuyama1978, DiCarlo1994}, where the CDW phase distortion is distributed over a spatial range across the pinning site\cite{Fukuyama1978}. However the short correlation length ($\sim 8$ unit cells) observed at $p=0.62$ GPa is unlikely to accommodate several disorder sites within a coherent volume to pin the CDW domain\cite{Feng2015}. Thus, we believe that the mechanism for destroying the long-range CDW modulation in ${\rm Ca_3Ir_4Sn_{13}}$ is probably the increasing quantum fluctuations when $T \rightarrow 0$ K and $p \rightarrow p_c$. Indeed, muon spin relaxation measurements\cite{Biswas15} have pointed out the importance of fluctuations when approaching the quantum critical point, which might be the origin of the enhancement of the superconducting phase above $p_c$. Quantum fluctuations has also been highlighted in LuPt$_{2-x}$Pd$_x$In system, where $T_C$ presents a dome-shaped doping dependence with highest value exactly where the CDW transition disappears\cite{Gruner2017}. Moreover, recent study of the CDW order parameter critical exponent in cubic intermetallics, including the (Ca$_{1-x}$Sr$_x$)$_3$Ir$_4$Sn$_{13}$ family of compounds, reveals a crossover of the classical thermal-driven CDW order parameter critical exponent expected for a three-dimensional universality class ($\beta \approx$0.3) to a mean-field tendency ($\beta \approx$0.5) as $T^* \rightarrow 0$\cite{carneiro2019}. This mean-field-like phenomenology supports the increase of dimensionality due to quantum fluctuations and provides evidence for the existence of a QCP in these compounds.


Finally, we have constructed the temperature-pressure phase diagram of Fig.~\hyperref[fig:Fig4]{5}. Results from XRD measurements on ${\rm (Sr,Ca)_3Ir_4Sn_{13}}$, electrical resistivity experiments on Ca$_3$Ir$_4$Sn$_{13}$ as well as other studies found in literature are reported. The phase diagram also depicts the apparent short-range order phase related to the partial suppression of the superlattice peak intensity in Ca$_3$Ir$_4$Sn$_{13}$ and possibly to the quantum fluctuations in this material. To better follow the evolution of such phase, we have extracted the temperature $T'$ from the maximum of the first derivative of the temperature dependence of the superlattice peak intensity for $T<15$ K (see Fig. S6 of the Supplemental Material~\cite{Note1} for more details). As mentioned before, our results suggests an enhancement of the quantum fluctuations when $T \rightarrow 0$ K and $p \rightarrow p_c$, which is likely the mechanism behind the partial/total suppression of the superlattice modulation at low temperatures. Although not probed in our experiment, we believe that a reentrant CDW modulation at higher pressures are unlikely to happen due to the lack of observation of such feature in bulk measurements. The presence of quantum fluctuations competing with CDW modulation and possibly with superconductivity makes the phase diagram of ${\rm (Sr,Ca)_3Ir_4Sn_{13}}$ reminiscent of unconventional superconductors. Indeed, it has been reported that these systems also show some unusual properties similar to the Fe-pnictides high temperature superconductors~\cite{Luo2018}, providing further evidence of the rich phase diagrams displayed by these materials. Particularly for ${\rm Ca_3Ir_4Sn_{13}}$, the different type of orders occur on comparable temperature scales and can compete/cooperate on an almost equal footing revealing their intertwined nature. Further experimental efforts will help to determine more accurately the P-T phase diagram as well as the nature of the quantum fluctuations in the low-temperature, high-pressure phase. For instance, measurements of the diffuse elastic line width in the anomalous phase could determine the critical exponent that controls the correlation length of the fluctuations.



\section{Conclusion}

Here we have performed a detailed study of the evolution of the superlattice structure of $\rm{(Sr, Ca)_3Ir_4Sn_{13}}$ against pressure by means of high-energy XRD measurements. We found that the superlattice transition temperature $T^*$ is rapidly suppressed with increasing pressure and extrapolates to zero at a critical pressure of $p_c \sim 1.79(4)$ GPa. Our XRD measurements on ${\rm Ca_3Ir_4Sn_{13}}$ revealed an anomaly related to a partial suppression of the superlattice peak intensity, which takes place at low temperatures ($T<15$ K) and high pressures ($p>0.09$ GPa). Such anomaly is also manifested by an increase of the pseudo-Voigt linewidth of the 2$\theta$ scans when the temperature approaches to zero. With no apparent origin on a structural phase transition or a competition with the superconducting phase that emerges at $T_C \sim 7$ K and reaches its maximum at $p\sim4$ GPa, our results suggest that quantum fluctuation effects is possibly the mechanism behind the destruction of the long-range CDW modulation. The revisited temperature-pressure phase diagram of $\rm{(Sr, Ca)_3Ir_4Sn_{13}}$ highlights the intertwined nature of the distinct order parameters and demonstrates some similarities of this family of supposedly conventional BCS superconductors~\cite{Biswas15, Gerber2013, Biswas2014, Biswas2014_2, Kase2011} and the unconventional superconductors.




\section*{acknowledgments}
We thank O. Gutowski for his assistance at P07 beamline of Petra III. We also thank M. Eleoterio, J. Fonseca and N. M. Souza-Neto for the help with high pressure diffraction measurements at XDS beam line at LNLS (proposal No. 20160202). Part of this research was carried out at PETRA III at DESY, a member of Helmholtz Association (HGF). LSI Veiga is supported by the UK Engineering and Physical Sciences Research Council (Grants No. EP/N027671/1 and No. EP/N034694/1). EM Bittar is supported by the Conselho Nacional de Desenvolvimento Cient\'ifico e Tecnol\'ogico (CNPq) (Grant No. 400633/2016-7) and Funda\c{c}\~{a}o Carlos Chagas Filho de Amparo \`{a} Pesquisa do Estado do Rio de Janeiro (FAPERJ) [Grant No. E-26/202.798/2019]. Part of this research was supported by the Funda\c{c}\~{a}o de Amparo \`{a} Pesquisa do Estado de S\~{a}o Paulo (FAPESP) (Grants No. 2012/04870-7 465 and No. 2017/10581-1).

\bibliography{references}

\end{document}


\title{Supplemental Material to: Possible quantum fluctuations in the vicinity of the quantum critical point of $\mathbf{(Sr, Ca)_3Ir_4Sn_{13}}$ revealed by high-energy X-ray diffraction study  }

\author{L.S.I. Veiga}

\affiliation{Deutsches Elektronen-Synchrotron (DESY), Hamburg 22607, Germany}

\affiliation{London Centre for Nanotechnology and Department of Physics and Astronomy, University College London, Gower Street, London, WC1E 6BT, United Kingdom}

\author{J.R.L. Mardegan}

\affiliation{Deutsches Elektronen-Synchrotron (DESY), Hamburg 22607, Germany}

\author{M.v Zimmermann}

\affiliation{Deutsches Elektronen-Synchrotron (DESY), Hamburg 22607, Germany}

\author{D.T. Maimone}
\affiliation{Instituto de F\'isica "Gleb Wataghin", Universidade Estadual de Campinas-UNICAMP, Campinas, S\~ao Paulo 13083-859, Brazil}

\author{F. B. Carneiro}
\affiliation{Instituto de F\'isica, Universidade do Estado do Rio de Janeiro, 20550-900, Rio de Janeiro, RJ, Brazil}
\affiliation{Centro Brasileiro de Pesquisas F\'isica, 22290-180, Rio de Janeiro, RJ, Brazil}

\author{M. B. Fontes}
\affiliation{Centro Brasileiro de Pesquisas F\'isica, 22290-180, Rio de Janeiro, RJ, Brazil}

\author{J. Strempfer}
\affiliation{Deutsches Elektronen-Synchrotron (DESY), Hamburg 22607, Germany}

\author{E. Granado}
\affiliation{Instituto de F\'isica "Gleb Wataghin", Universidade Estadual de Campinas-UNICAMP, Campinas, S\~ao Paulo 13083-859, Brazil}

\author{P. G. Pagliuso}
\affiliation{Instituto de F\'isica "Gleb Wataghin", Universidade Estadual de Campinas-UNICAMP, Campinas, S\~ao Paulo 13083-859, Brazil}

\author{E.M. Bittar}
\affiliation{Centro Brasileiro de Pesquisas F\'isica, 22290-180, Rio de Janeiro, RJ, Brazil}

\maketitle

\beginsupplement

\section{Raw XRD data at $p=0.7$ GPa}

\begin{figure}[h]
	\centering
	\includegraphics[width=1.\columnwidth]{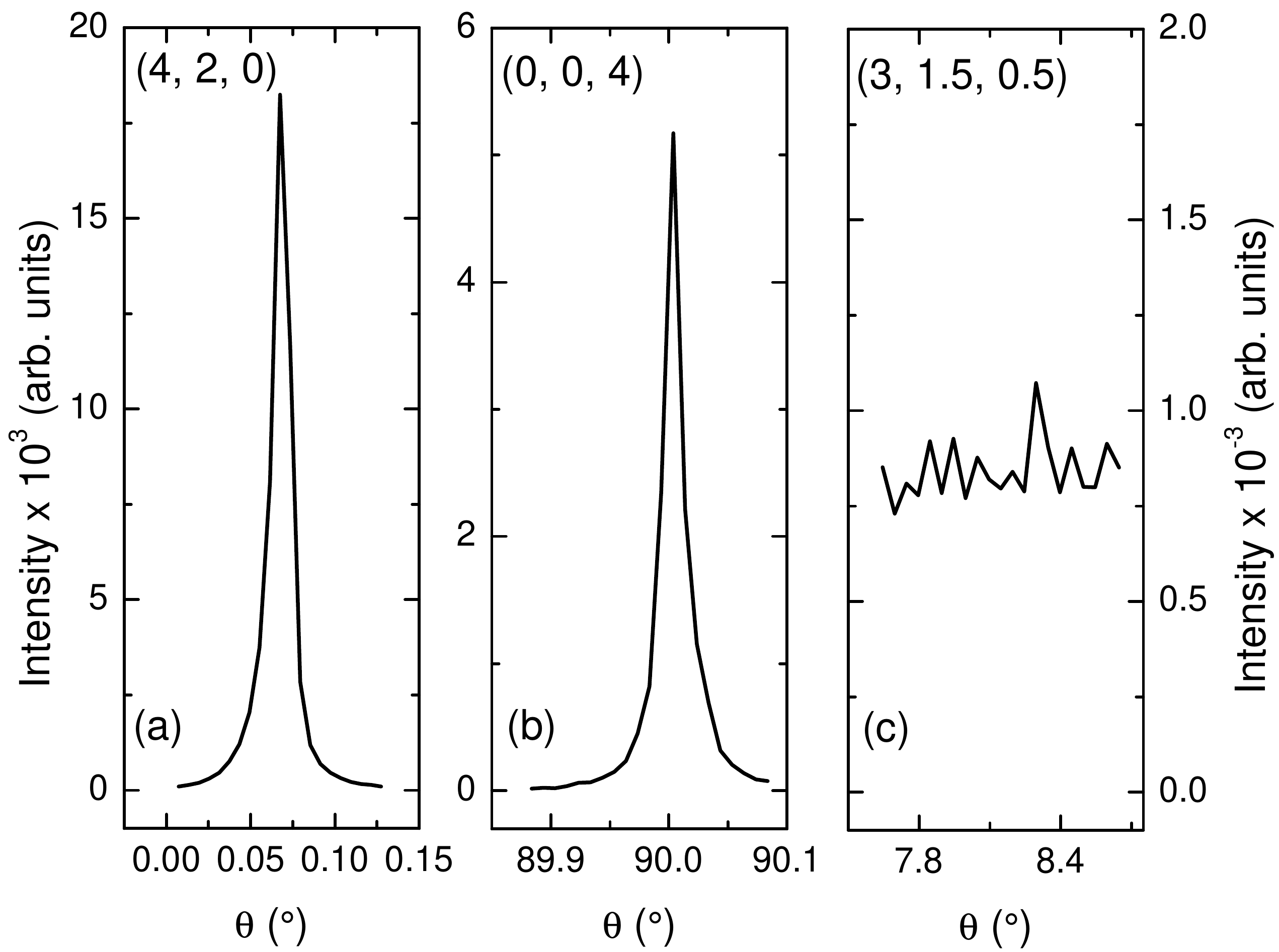}

	\caption{ Rocking curves at the lattice Bragg reflections (a) $(4,2,0)$ and (b) $(0,0,4)$ and superlattice Bragg reflection (c) $(3,1.5,0.5)$ at P = 0.70 GPa and T = 5 K.} 
	\label{fig:FigS1}

\end{figure}

The rocking curve scans at the lattice Bragg reflections $(4,2,0)$, $(0,0,4)$ and superlattice Bragg peak $(3,1.5,0.5)$ are shown in Figure~\hyperref[fig:FigS1]{S1}. These data were collected right after the collection of the P = 0.62 GPa data set, i. e., the next pressure point P=0.7 GPa. While the rocking curves of the lattice Bragg peaks $(4,2,0)$ and $(0,0,4)$ are still present at this pressure, we could not find the superlattice reflection at the calculated position determined by the UB matrix. In addition, no superlattice peak intensity was found by extending the angular limits of the rocking scans, changing temperature (from 3 to 15 K) or by searching other superlattice reflections respecting the wave vector $\mathbf{q_{\rm SL}}=(0.5, 0.5, 0)$. This suggests the total suppression of the CDW modulation intensity was reached. 

\section{Pressure determination from the orthorhombic splitting of (200)/(020) Bragg peaks in $\mathbf{La_{1.875}Ba_{0.125}CuO_4}$}
\begin{figure}[t]
	\centering
	\includegraphics[width=0.6\columnwidth]{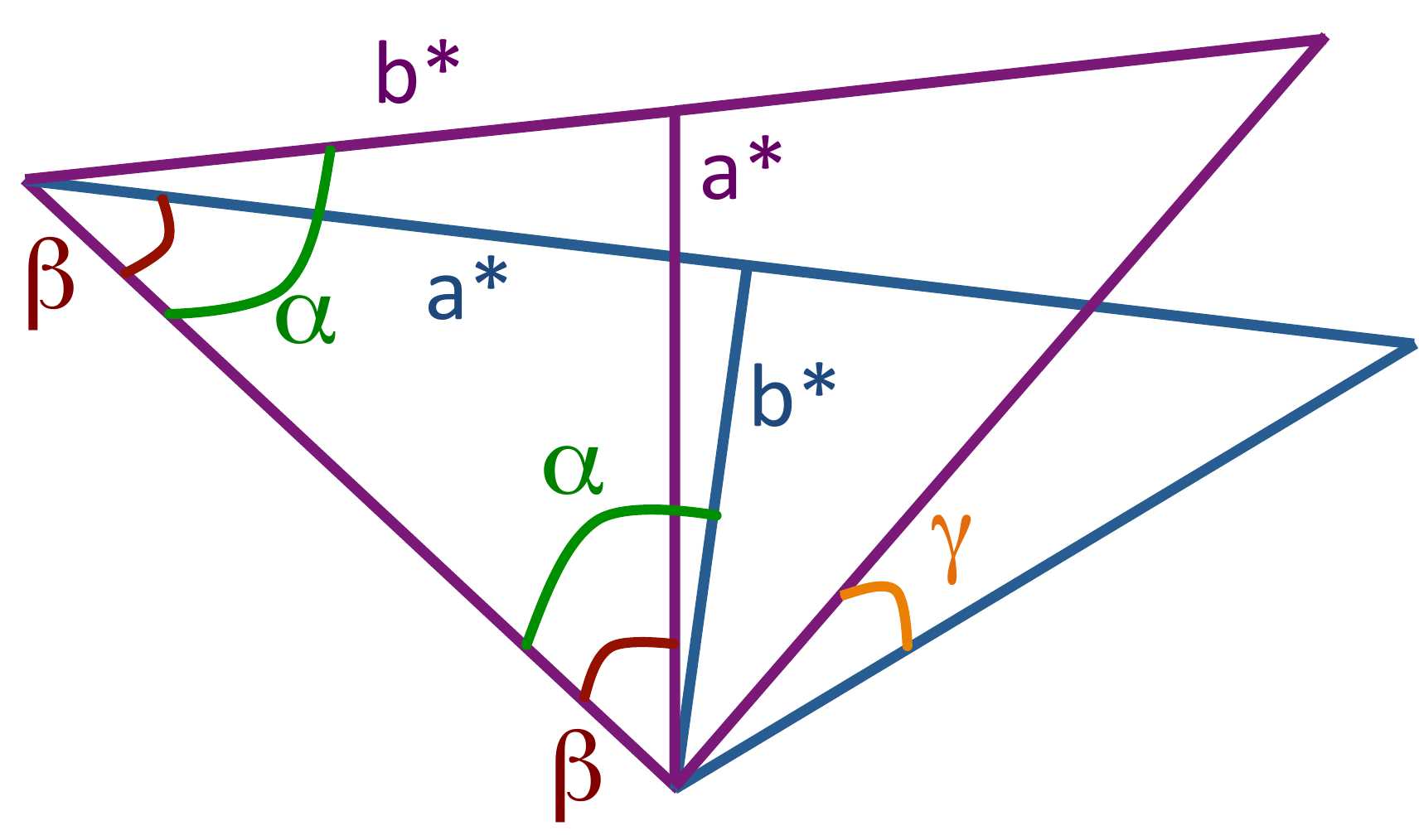}

	\caption{ Schematic illustration of the orthorhombic splitting of the $(2,0,0)/(0,2,0)$ Bragg peaks in La$_{1.875}$Ba$_{0.125}$CuO$_4$.}
	\label{fig:twinning}

\end{figure}      

\begin{figure}[t]
	\centering
	\includegraphics[width=0.6\columnwidth]{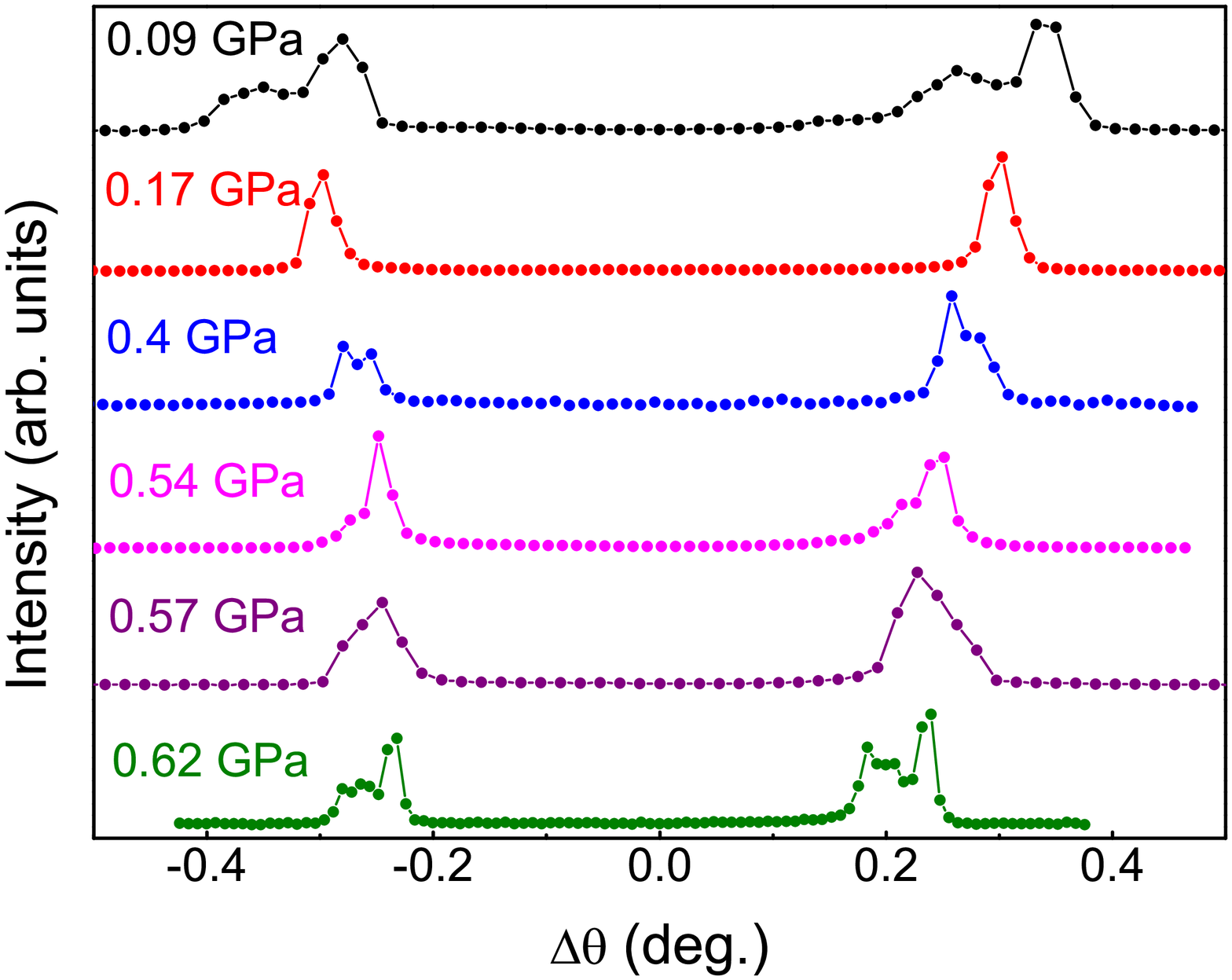}

	\caption{ Rocking scans ($\theta$ scans) through the $(2,0,0)$ and $(0,2,0)$ Bragg reflections, showing the evolution of the orthorhombic splitting with pressure.}
	\label{fig:splitting}

\end{figure}

At ambient pressure, La$_{1.875}$Ba$_{0.125}$CuO$_4$ compound shows two structural transitions: from the high-temperature-tetragonal (HTT) phase to the low-temperature-orthorhombic (LTO) phase at $T_{HTT}\approx 235$ K, and from the LTO phase to the low-temperature-tetragonal (LTT) phase at $T_{LTO}\approx54$ K~\cite{Huecker10}. In the orthorhombic phase, La$_{1.875}$Ba$_{0.125}$CuO$_4$ form twin domains, from which the orthorhombic strain can be determined by means of transverse scans through the pair of $(2,0,0)/(0,2,0)$ Bragg reflections. We have explored the pressure dependence of this splitting over a range of temperature ($T\sim60-66$ K). Fig.~\ref{fig:splitting} presents the corresponding scans at various pressures. Pressure calibration is made via the orthorhombic strain $s=2(b-a)/(a+b)$, where $a$ and $b$ are the in-plane lattice constants. More specifically, this quantity can be determined from the sets of equations below (see Fig.~\ref{fig:twinning}):

\begin{equation}
\begin{array}{cc}
\displaystyle \alpha = arctan\frac{a^*}{b^*} = arctan\frac{b}{a}  \\\\
\displaystyle \beta = arctan\frac{b^*}{a^*} = arctan\frac{a}{b}
\end{array}
\label{equation1}
\end{equation}

\noindent with $a^*>b^*$ in the reciprocal space and $b>a$ in real space. $\gamma$ is the angle separating the $(2,0,0)/(0,2,0)$ Bragg reflections and is given by $\gamma = 2\alpha-2\beta$, so that:

\begin{equation}
\frac{\gamma}{2} = arctan\frac{b}{a}-arctan\frac{a}{b} = arctan\left( \frac{b^2-a^2}{2ab}\right)
\label{equation2}
\end{equation}

\noindent Considering that $a \approx b$, then:

\begin{equation}
\begin{array}{cc}
\displaystyle \frac{a}{b} = \sqrt{1-2tan\left( \frac{\gamma}{2}\right)} \\\\
\displaystyle \frac{b}{a}-1\approx 2\frac{b-a}{a+b}=s
\end{array}
\label{equation3}
\end{equation}

Pressure can be calibrated when $s(p)$ is known for a certain doping ($x$) and temperature ($T$), through the equation:
  
\begin{equation}
\xi = 1-\frac{T^*}{T_0}-\frac{p}{p_0}-\frac{x}{x_0}
\label{equation4}
\end{equation}

\noindent where $s$ is a linear function of $\xi$ as observed in Refs~\onlinecite{vonZimmerman08, Takahashi1994}. $T^*=\Theta$ ${\rm coth}(\Theta/T)$, $T_0$ is the  temperature of the critical high-temperature structural transition $HTT \leftrightarrow LTO$ in La$_2$CuO$_4$ ($x=0$), $\Theta$ accounts for the nonlinearity of $s(T)$ at low T. $p_0$ is the critical pressure in La$_2$CuO$_4$, and $x_0$ is the critical Ba concentration in La$_{2-x}$Ba$_x$CuO$_4$ to suppress the orthorhombic phase. All these parameters were extracted from Ref.~\onlinecite{Takahashi1994}.

\section{Pseudo-Voigt linewidth analysis of the $2\theta$ scans of Ca$_3$Ir$_4$Sn$_{13}$, correlation length calculation and first derivative of the Intensity vs Temperature curves}

\begin{figure}[t]
	\centering
	\includegraphics[width=0.7\columnwidth]{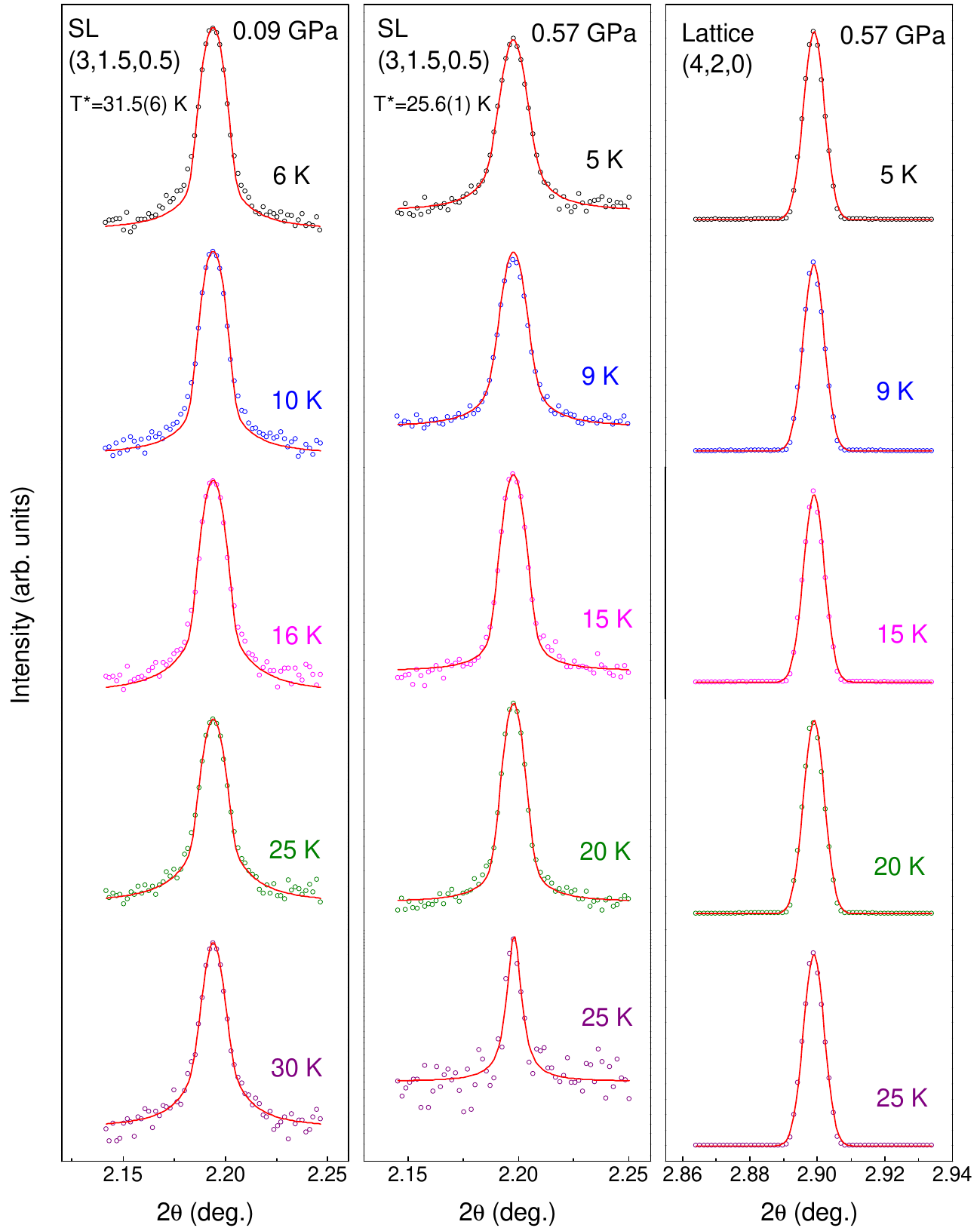}

	\caption{ Temperature evolution of the lattice and superlattice diffraction line shapes for $p=0.09$ GPa and 0.57 GPa.} 
	\label{fig:evolution}

\end{figure}

\begin{figure}[t]
	\centering
	\includegraphics[width=0.6\columnwidth]{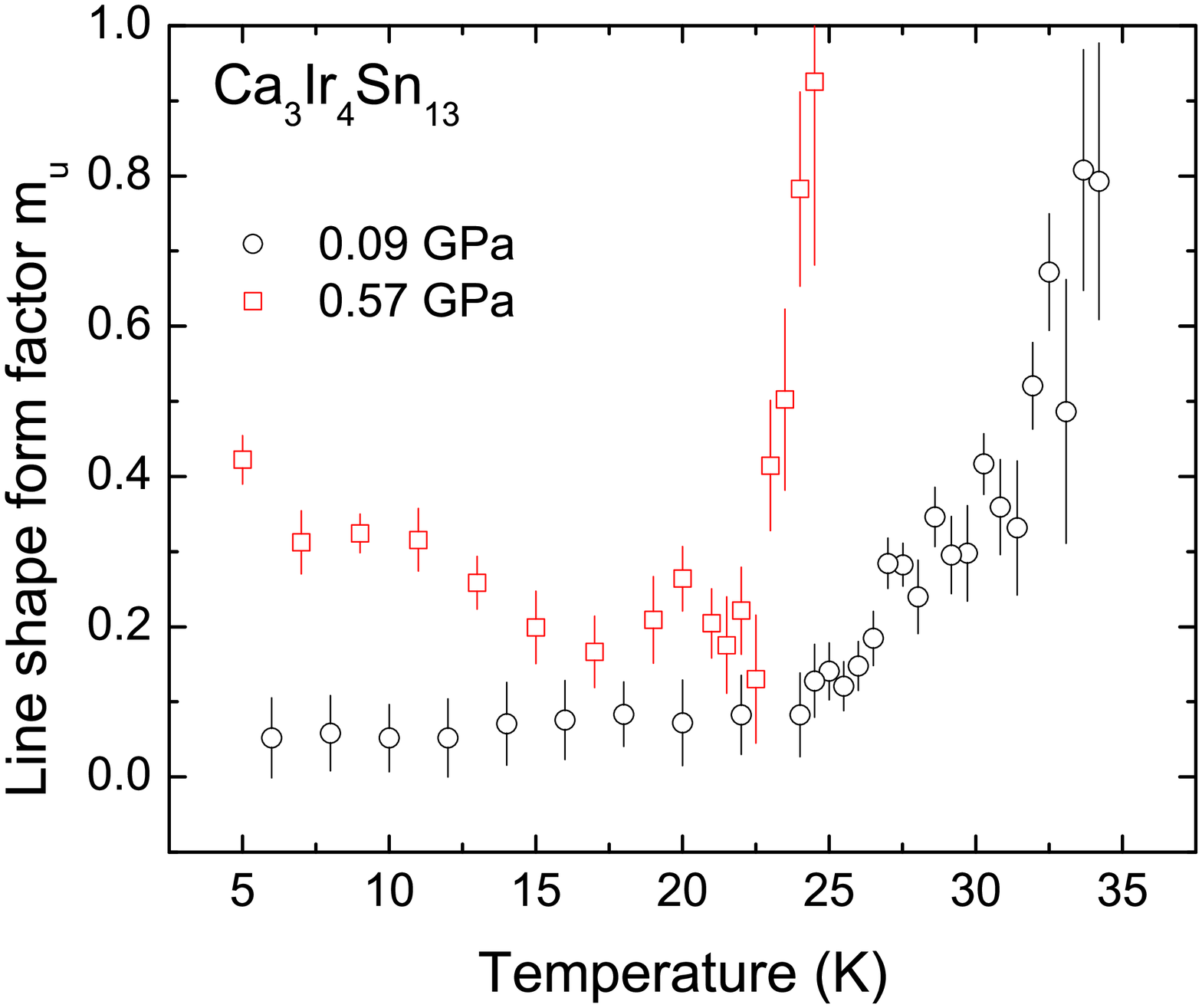}

	\caption{Spectral line shape form factor $m_u$ extracted from the fitting of the superlattice $2\theta$ scans by a pseudo-Voigt function (Equation~\ref{Eq1}). The spectral shape assumes a pure Lorentzian form when $m_u=1$ and a pure Gaussian character when $m_u=0$.}
	\label{fig:form_factor}

\end{figure}

The $2\theta$ scans of the $(3,1.5,0.5)$ superlattice reflection were fitted using a pseudo-Voigt spectral line shape of the form:

\begin{equation}
y=y_0+A\left[m_u\frac{2}{\pi}\frac{w}{4(x-x_c)^2+w^2}+(1-m_u)\frac{\sqrt{4ln2}}{\sqrt{\pi}w}e^{-\frac{4ln2}{w^2}(x-x_c)^2}\right]
\label{Eq1}
\end{equation}

\noindent which is a linear combination of a Gaussian and Lorentzian functions. Here, $y_0$ is the offset, $x_c$ the peak center. The quantities of interest in the analysis of the superlattice peaks are the integrated area, $A$, the peak width, $w$ and the profile shape factor, $m_u$. Figure~\ref{fig:evolution} shows the temperature evolution of the lattice and superlattice reflections at selected pressures. The line width of the $(4,2,0)$ lattice peak at $p=0.57$ GPa is essentially constant as a function of temperature, as observed in Fig.~\ref{fig:Fig1} of the main text. The pseudo-Voigt line shape indicates a comparable contribution from the sample lattice and instrumental resolution. At $p=0.09$ GPa, the superlattice peaks broadens quickly when $T \rightarrow T^*$, and the spectral line shape changes from a Gaussian character at low temperatures to a Lorentzian form. At higher pressures, on the other hand, the profile shape has a significant contribution from the Lorentzian line shape already at low temperatures, as seen in Fig.~\ref{fig:form_factor}. The Lorentzian line shape provides direct evidence of quantum fluctuations when $T\rightarrow0$ K and $p \rightarrow p_c$ and is likely the mechanism behind the partial/total suppression of the CDW modulation.

The static correlation length was extracted from the $2\theta$ scans by the equation, $\xi=1/w$, where $w=\sqrt{w_{meas}^2-w_R^2}$ is the measured pseudo-Voigt line width corrected for the instrument resolution $w_R$ and expressed in units of ${\AA}^{-1}$. The instrumental resolution was extracted from the line width of the lattice Bragg reflections, which is essentially pure Gaussian line shapes.

\begin{figure}
	\centering
	\includegraphics[width=0.5\columnwidth]{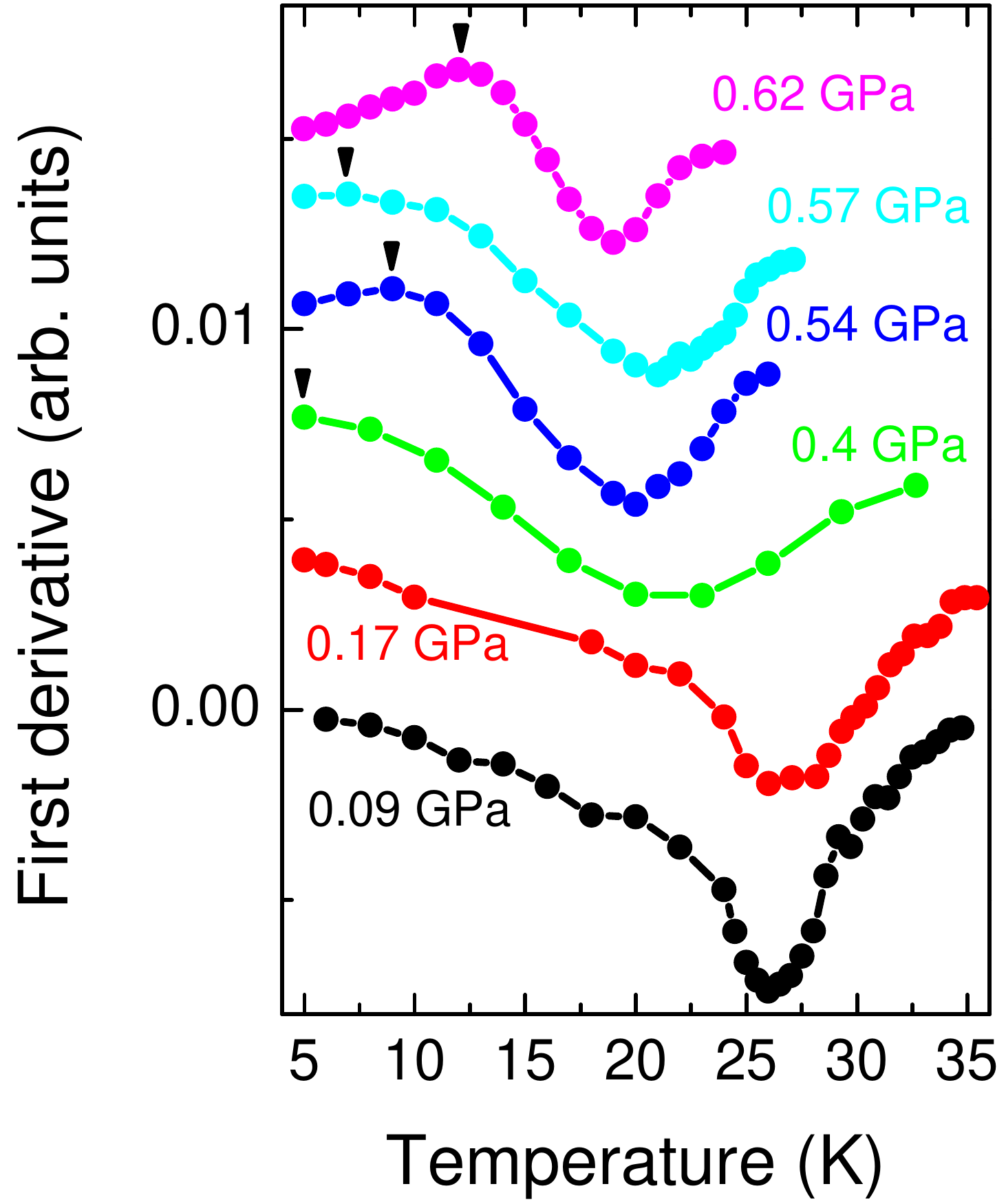}

	\caption{Evolution of the maximum of the first derivative (indicated by arrows) of the superlattice peak intensity with respect to temperature for several pressures.}
	\label{fig:first_derivative}

\end{figure}

The phase diagram shown in Fig.~\ref{fig:Fig4} of the main manuscript depicts the apparent short-range order phase which is possibly associated with the quantum fluctuations in Ca$_3$Ir$_4$Sn$_{13}$ when $T \rightarrow 0$ K and $p \rightarrow p_c$ . Such phase is manifested by the partial suppression of the superlattice Bragg peak intensity at low temperatures observed in our XRD measurements. We investigate the evolution of such phase by following the maximum of the first derivative ($T'$) of the temperature dependence of the superlattice peak intensity for $T<15$ K as shown in Fig.~\ref{fig:first_derivative}. Other methods to extract $T'$, such as fitting the curve using  the powder law $\propto (1-T/T^*)^{2\beta}$, were not possible since the total suppression of the superlattice Bragg peak was not reached even for the highest pressure achieved in this experiment ($P=0.62$ GPa).

\section{Resistivity measurements on Ca$_3$Ir$_4$Sn$_{13}$}

Temperature dependent electrical resistance measurements were carried out via a conventional four-point method driven by a LR-700 AC Resistance Bridge in a closed cycle cryostat.

\begin{figure}
	\centering
	\includegraphics[width=0.6\columnwidth]{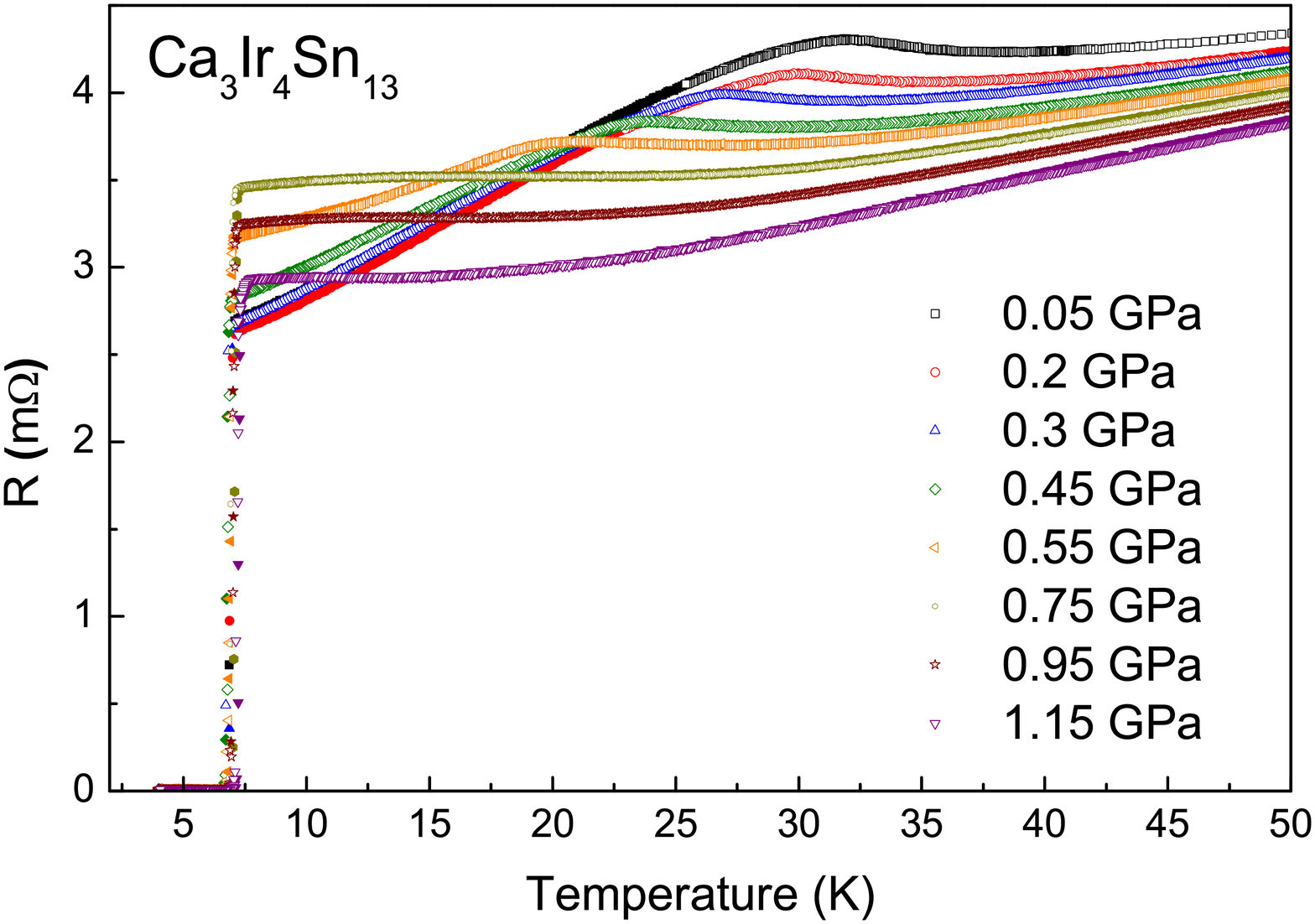}

	\caption{Representative pressure-dependent electrical resistance curves as a function of temperature of the superlattice structural transition $T^*$ and superconducting transition $T_C$ in Ca$_3$Ir$_4$Sn$_{13}$.} 
	\label{fig:resistivity}

\end{figure}

The temperature dependent electrical resistance $R(T)$ of Ca$_3$Ir$_4$Sn$_{13}$ single crystals, at ambient pressure, show metallic behavior down to low temperatures, until the superconducting transition takes place at $T_C\sim7$ K. A small kink due to the superlattice structural transition is observed at $T^*\sim37$ K, consistent with our synchrotron XRD data. No thermal hysteresis was observed for $R(T)$ in the entire pressure range measured. Representative curves of the pressure evolution of $T^*$ and $T_C$ in Ca$_3$Ir$_4$Sn$_{13}$ for our $R(T )$ data are presented in Fig.~\hyperref[fig:resistivity]{S7}. The estimated pressure error bar is $\pm0.05$ GPa.

\clearpage

\bibliography{references}